\documentclass[conference]{IEEEtran}
\IEEEoverridecommandlockouts
\usepackage{cite}
\usepackage{amsmath,amssymb,amsfonts}
\usepackage{graphicx}
\usepackage{textcomp}
\usepackage{xcolor}
\usepackage{float}
\usepackage{pifont}

\usepackage{graphicx}
\usepackage{xcolor}
\usepackage{color,soul}
\usepackage[caption=false]{subfig}
\usepackage{amssymb}
\usepackage{amsmath}
\usepackage{rotating}
\usepackage{enumerate}
\usepackage{enumitem}

\usepackage{textcomp}
\usepackage{calrsfs}
\usepackage{float}
\usepackage{listings}
\usepackage{csquotes}
\usepackage[switch]{lineno}
\usepackage{enumerate}
\usepackage[caption=false]{subfig}
\def\BibTeX{{\rm B\kern-.05em{\sc i\kern-.025em b}\kern-.08em
    T\kern-.1667em\lower.7ex\hbox{E}\kern-.125emX}}
\begin{document}

\title{An Annexure to the Paper - ``Driving the Technology Value Stream by Analyzing App Reviews"}

\author{Souvick~Das,~\IEEEmembership{}
Novarun~Deb,~\IEEEmembership{Member,~IEEE,}
        Nabendu~Chaki,~\IEEEmembership{Senior Member,~IEEE,}
        and~Agostino~Cortesi~\IEEEmembership{}
\thanks{S. Das and A. Cortesi are with the DAIS Department, Ca' Foscari University, Venice, Italy. \texttt{e-mail}: \textit{souvik.cmsa019@gmail.com,cortesi@unive.it}}
\thanks{N. Chaki is with the Department
of Computer Science and Engineering, University of Calcutta, India. \texttt{e-mail}: \textit{nabendu@ieee.org}} 
\thanks{N. Deb is with the Indian Institute of Information Technology (IIIT), Vadodara, India.  \texttt{e-mail}: \textit{novarun\_deb@iiitvadodara.ac.in}}}

\maketitle


\section{Introduction}
The paper titled ``Driving the Technology Value Stream by Analyzing App Reviews" presents a novel framework that utilizes Natural Language Processing (NLP) techniques to understand user feedback on mobile applications. The framework allows software companies to drive their technology value stream based on user reviews, which can highlight areas for improvement. The framework is analyzed in depth, and its modules are evaluated for their effectiveness. The proposed approach is demonstrated to be effective through an analysis of reviews for sixteen popular Android Play Store applications over a long period of time.

This supplementary document complements the proposed framework for analyzing application reviews and comprises a thorough description of the deep learning models utilized in the framework. Furthermore, we present an analysis of the experimental outcomes of the framework. The document concludes by providing an evaluation of the effectiveness of the framework.
\section{Deep Learning Approaches Used in the Framework}
This section provides a detailed explanation of various deep-learning techniques used in the proposed framework of the paper titled ``Driving the Technology Value Stream by Analyzing App Reviews".
\subsection{Prompt Engineering Based Zero-shot Classifier}
\textit{\textbf{What is zero-shot learning?}} According to a study~\cite{zsl}, an approach to scaling up the recognition task by developing such models capable of recognizing unseen classes without any prior training is called zero-shot learning.
\par \textit{\textbf{Why is it emerging?}} Large-scale annotated dataset availability is necessary for deep learning techniques. The major obstacle to completing down-streaming tasks in the field of App review analysis is the unavailability of such publicly accessible datasets. 
\par \textit{\textbf{How to rescue?}} In rescue, Zero-Shot Learning is the solution where it uses deep learning models to classify examples from new classes for which there is no need for prior training data. 
\par \textit{\textbf{How to achieve?}} An innovative zero-shot learning method called prompt-based learning involves creating templates for different kinds of NLP tasks.
\subsubsection{ Prompt Learning and Template Design} One of the latest approach to achieve zero-shot learning is prompt-based learning which involves developing templates for specific NLP tasks. The process of building prompting functions or templates depending upon the input text data and the specific task is called prompt engineering. Traditional supervised learning trains a model to take an input \textit{\textbf{x}} and predict an output \textit{\textbf{y}} denoted as \textit{\textbf{P(y}} $\rvert$ \textbf{\textit{x)}}.
On the other hand, prompt-based learning is based on language models that model the probability of text directly~\cite{zeroshot}. The template is developed on top of the original text \textit{\textbf{x}} and usually, the template has some unfilled slots.
The language models perform prediction tasks on the template and predict to fill the unfilled slots and generate a string $\hat{\textit{\textbf{x}}}$. The final output \textit{\textbf{y}} is then derived from $\hat{\textit{\textbf{x}}}$. 
\par In our work, we take the user reviews as the input text and define templates including one mask token as an unfilled output slot. Finally, we choose a language model to predict the highest scoring $\hat{\textit{\textbf{y}}}$ for the output label. We elaborate on each step with examples for better understandability.
\subsubsection{ Prompt Creation}
\begin{table}[htb]
\caption{Notations and steps involved in prompt learning.}
\resizebox{.5\textwidth}{!}{\begin{tabular}{lcll}
\multicolumn{1}{c}{\textbf{Name}}                          & \textbf{Notation}                     & \multicolumn{1}{c}{\textbf{Example}}                                                                                  & \multicolumn{1}{c}{\textbf{Explanation}}                                                                                             \\ \hline
Input                                                      & \textit{\textbf{x}}                   & \begin{tabular}[c]{@{}l@{}}Virtual background\\ of the App is not working.\end{tabular}                               & Input text                                                                                                                           \\
Output                                                     & \textit{\textbf{y}}                   & Fault                                                                                                                 & Output label                                                                                                                         \\
\begin{tabular}[c]{@{}l@{}}Possible\\ Labels\end{tabular}  & \textit{\textbf{z}}                   & \begin{tabular}[c]{@{}l@{}}\{Bug, Fault, Praise, Feature\\ request, ...{]}\end{tabular}                               & \begin{tabular}[c]{@{}l@{}}Set of labels from which the\\ output will be decided.\end{tabular}                                       \\ \hline
\begin{tabular}[c]{@{}l@{}}Prompt \\ function\end{tabular} & \multicolumn{1}{l}{$f_{prompt} (x)$} & {[}X{]}. The review is about {[}Z{]}                                                                                  & \begin{tabular}[c]{@{}l@{}}A function creates a template\\ from input and adds a slot Z\\ that will be filled later.\end{tabular} \\ \hline
Prompt                                                     & $\acute{x}$                                   & \begin{tabular}[c]{@{}l@{}}Virtual background\\ of the App is not working.\\ The review is about {[}Z{]}\end{tabular} & \begin{tabular}[c]{@{}l@{}}The modified text where {[}X{]}\\ is filled but {[}Z{]} is not.\end{tabular}                              \\
\begin{tabular}[c]{@{}l@{}}Filled\\ Prompt\end{tabular}    & \multicolumn{1}{l}{$f_{fill} (\acute{x}, z^*)$}     & \begin{tabular}[c]{@{}l@{}}Virtual background\\ of the App is not working.\\ The review is about praise.\end{tabular} & \begin{tabular}[c]{@{}l@{}}{[}Z{]} can be filled with any \\ answer.\end{tabular}                                                    \\
\begin{tabular}[c]{@{}l@{}}Answer\\ Prompt\end{tabular}    & \multicolumn{1}{l}{$f_{fill} (\acute{x}, z)$}   & \begin{tabular}[c]{@{}l@{}}Virtual background\\ of the App is not working.\\ The review is about fault.\end{tabular}  & {[}Z{]} is filled with true answer.                                                                                                  \\ \hline
\end{tabular}}
\label{prompt_notation}
\end{table}
In this stage, \textit{prompting function} $f_{prompt} (x)$ is applied on input \textbf{\textit{x}} and transform the input into $\acute{x}$. Thus $\acute{x} = f_{prompt} (x)$. At first, we define a template, that includes two slots - an input slot [X] for input \textbf{\textit{x}} and an answer slot [Z]. The slot [X] is filled by the input text \textbf{\textit{x}}. The answer slot can be filled with any labels as the intermediate answer text. Later, the highest scored predicted label will be mapped into \textbf{\textit{y}}. In table \ref{prompt_notation}, we present notations and steps involved in prompt learning with corresponding examples. There are two types of prompt possible in prompt engineering. In table \ref{prompt_types}, we show examples for each of the categories of prompt template. In \textit{Prefix prompt}, the tokens predicted by a language model fill the subsequent masks. In case of \textit{Cloze prompt}, the slots are in somewhere middle of the string and it can be viewed as fill in the blank query. 

\begin{table}[htb]
\caption{Types of Prompt Template}
\begin{tabular}{|l|l|}
\hline
\textbf{Prompt Template}                                 & \textbf{Type} \\ \hline
{[}X{]}. The review is about {[}Z{]}                     & Prefix Prompt \\ \hline
{[}X{]} : The review specifies {[}Z{]} type of category. & Cloze Prompt  \\ \hline
\end{tabular}
\label{prompt_types}
\end{table}
\subsubsection{ Text Entailment Based Zero-shot  Classifier}
Our work utilizes the text entailment approach to create a zero-shot classifier for the task of topic classification. This process involves three steps. Firstly, we address the textual entailment problem. In second step, we design templates that can solve the textual entailment problem while also generating the zero-shot classifier. Finally, in the last step, a language model is used to solve the text entailment problem and provides the classification result. 
\begin{table}[htb]
\caption{Prompt Engineering for proposed zero-shot classifier}
\resizebox{.5\textwidth}{!}{\begin{tabular}{lll}
\textbf{Name}                                                              & \textbf{Example}                                                                                                                                                                           & \textbf{Explanation}                                                                                                                      \\ \hline
Input (X1)                                                                 & \begin{tabular}[c]{@{}l@{}}The virtual background of \\ the App is not working.\end{tabular}                                                                                               & The input review statement                                                                                                                \\
Input (X2)                                                                 & Sentiment: Negative                                                                                                                                                                        & Another input as sentiment                                                                                                                \\
\begin{tabular}[c]{@{}l@{}}Possible labels \\         (Z)\end{tabular}     & \begin{tabular}[c]{@{}l@{}}\{Bug, Feature request, Fault,\\  Praise, ..... \} and def(\{Bug, \\  Feature request, ...\})\end{tabular}                                                                                                 & \begin{tabular}[c]{@{}l@{}}Z is the set of candidate labels\\ along with their definitions\end{tabular}                                                       \\ \hline
\begin{tabular}[c]{@{}l@{}}Prompt function\\     ($f_{prompt} (x)$)\end{tabular} & \begin{tabular}[c]{@{}l@{}}{[}X1{]}. Sentiment of the \\ review is {[}X2{]}. The review \\ is about {[}Z{]}.\end{tabular}                                                                  & \begin{tabular}[c]{@{}l@{}}A function creates a template \\ from input and adds a slot {[}Z{]} \\ that will be filled later.\end{tabular} \\ \hline
Prompt ($\acute{x}$)                                                             & \begin{tabular}[c]{@{}l@{}}The virtual background of \\ the App is not working. \\ Sentiment of the review is \\ Negative. The review is about \\ {[}Z{]}.\end{tabular}                    & \begin{tabular}[c]{@{}l@{}}The modified text where {[}X1{]}\\ and {[}X2{]} are filled but not {[}Z{]}\end{tabular}                        \\
                                                                           &                                                                                                                                                                                            &                                                                                                                                           \\
\begin{tabular}[c]{@{}l@{}}Filled Prompt\\ ($f_{fill} (\acute{x}, z^*)$)\end{tabular}      & \begin{tabular}[c]{@{}l@{}}The virtual background of \\ the App is not working. \\ Sentiment of the review is \\ Negative. The review is about \\ bug / fault / praise / ....\end{tabular} & \begin{tabular}[c]{@{}l@{}}{[}Z{]} can be filled with any\\ labels.\end{tabular}                                                          \\
                                                                           &                                                                                                                                                                                            &                                                                                                                                           \\
\begin{tabular}[c]{@{}l@{}}Answer Prompt\\ ($f_{fill} (\acute{x}, z)$)\end{tabular}        & \begin{tabular}[c]{@{}l@{}}The virtual background of \\ the App is not working. \\ Sentiment of the review is \\ Negative. The review is about \\ bug i.e. a problem causing a \\ program to crash or produce \\ invalid outcome.\end{tabular}                        & {[}Z{]} is filled with true answer.                                                                                                       \\ \hline
\end{tabular}}
\label{our_template}
\end{table}
\begin{itemize}

    \item[-] \textit{Text Entailment:} It finds the relation between text fragments. The entailment relation holds between two text fragments if one text follows from another text holds true. This imitates how humans understand the meaning of an aspect and derive hypothesis from it. For example, if we consider the statement (S)- ``\textit{Arsenal sign striker Jesus from Manchester City}" and possible candidates \{\textit{Player contract signing, New manager appointment}\}. We humans generally,
    construct a hypothesis by taking a text like ``the text is about \_\_\_" and by filling the mask or blank space using candidates e.g., ``Player contract signing" or ``New manager appointment". Next, we try to evaluate whether this hypothesis is true for the given the statement S.
    
    \item [-] \textit{Defining Prompt Template:}
    In table \ref{our_template}, we have shown how we have defined the template for zero-shot classifier. The template definition has two input slots X1 and X2 where X1 is filled with the user review statement and X2 is to be filled by sentiment of the review. Thus the modified text ($\acute{x}$) becomes the premise. In order to create the hypotheses, we consider the labels and their definitions. We combine the labels and the definitions of the labels as well for generating different hypotheses. Recent study~\cite{zeroshot} suggests to use the combination of labels and definition as hypothesis to get better performance. We also found that, instead of selecting labels or definition individually for constructing hypothesis, the combination of labels and their definitions gives better performance. Based on the hypothesis and the premise the pre-trained BART-large-mnli model predict which particular hypothesis gets entailed with respect to the premise.

    \item [-] \textit{ Pre-training of Language Model}
    In order to predict the output label from the entailment task, a pre-trained language model is essential which is trained on natural language inference dataset. Thus, we go for an approach that uses a pre-trained MNLI sequence-pair classifier model as an out-of-the-box zero-shot text classifier. In this work, we train BART-large model on MNLI~\cite{mnli} dataset which performs with 54\% of f1-score for Yahoo Answers~\cite{yahoo} topic classification dataset. Yin et. al.~\cite{zeroshot} report a label-weighted f1-score of 37.9 on Yahoo Answers~\cite{yahoo} using the BERT model fine-tuned only on the MNLI dataset.
\end{itemize}
\subsection{ Practical Approach for Prompt-Based Zero-shot Classifier}
OpenPrompt~\cite{openprompt} library that provides provision to build the prompt learning pipeline. 
\subsubsection{ Language Model Selection}
OpenPrompt is compatible with pre-trained language models available on huggingface. At first, we train the BART-large model on MNLI and we uploaded the model in huggingface. In the next phase, we refer to our model during pre-training model selection phase in openprompt pipeline. Following code snippet shows how simple it is to load a language model.
\newline
\texttt{from openprompt.plms import load\_plm \newline
plm, tokenizer, model\_config, 
WrapperClass = \newline load\_plm("bert", 
            "/plm/bert-base-cased")}

\subsubsection{ Defining Template}
A Template modifies the original input text, which is also one of the most important modules in prompt-learning. The modification of text is carried out in order to make it more compatible and suitable for the particular task to be performed. For example consider the following text.\newline 

\texttt{from openprompt.prompts import ManualTemplate
\newline promptTemplate = ManualTemplate(
    \newline text = '\{"text\_a"\}. It \space is \{"mask"\} '
)}

In this example, where the $\langle$text\_a$\rangle$ will be replaced by the text from input data, and the $\langle$mask$\rangle$ will be used to predict a label word.

\subsubsection{ Defining Verbalizer}
A Verbalizer is another important (but not necessary such as in generation) in prompt-learning, which projects the original labels to a set of label words as shown in table \ref{tbl_verbalizer}. Basically, the language model tries to fill the masked word based on the context of the sentence and the existing knowledge it has. In order to use the verbalizer, we can define a dictionary of words for every label called label words. The pre-trained language model predicts a probability distribution over the vocabulary for one masked position. The verbalizer will extract the logits of label words and integrate the logits of label words to the corresponding class to widening the range for potential mapping with predicted word. Verbalizer can be defined in following manner.
\newline
\texttt{from openprompt.prompts import ManualVerbalizer
\newline promptVerbalizer = ManualVerbalizer(
    \newline label\_words = \{
        "bug": ["crash", "malfunction"],
        \newline "praise": ["good", "wonderful", "great"],
    \}
)}

\begin{table}[htb]
\caption{Verbalizer class with label words for different classes}
\begin{tabular}{|l|l|}
\hline
\textbf{Class}         & \textbf{Verbalizer class}                                                                       \\ \hline
Bug                    & crash, glitch, hang, malfunction                                                                \\ \hline
Fault                  & \begin{tabular}[c]{@{}l@{}}flaw, inconsistency, drawback, \\ failing, shortcomings\end{tabular} \\ \hline
Feature request        & \begin{tabular}[c]{@{}l@{}}functionality implementation,\\ new feature\end{tabular}             \\ \hline
Improvement suggestion & \begin{tabular}[c]{@{}l@{}}upgrade, update, enhancement, \\ modification\end{tabular}           \\ \hline
Information enquiry    & question, FAQ, help, support                                                                    \\ \hline
Content request        & \begin{tabular}[c]{@{}l@{}}text content addition, page \\ content\end{tabular}                  \\ \hline
Feature information    & about feature, highlights                                                                       \\ \hline
\end{tabular}
\label{tbl_verbalizer}
\end{table}
\par Once we define template and verbalizer, the entire pipeline is ready for conducting the training and inferencing for app review dataset using PyTorch\footnote{https://pytorch.org/docs/stable/index.html} library. The entire documentation of OpenPrompt can be found on their official documentation \footnote{https://thunlp.github.io/OpenPrompt/index.html}. We also provide our GitHub\footnote{https://github.com/svk-cu-nlp/zsl\_app\_review} repository containing entire source codes for zero-shot classifier that we have built. 

\subsection{The Key Phrase Extraction Mechanism}
\begin{figure}[htb]
\caption{Architecture of the Key Phrase Extraction}
	\centering
	\includegraphics[width=0.5\textwidth,page=1]{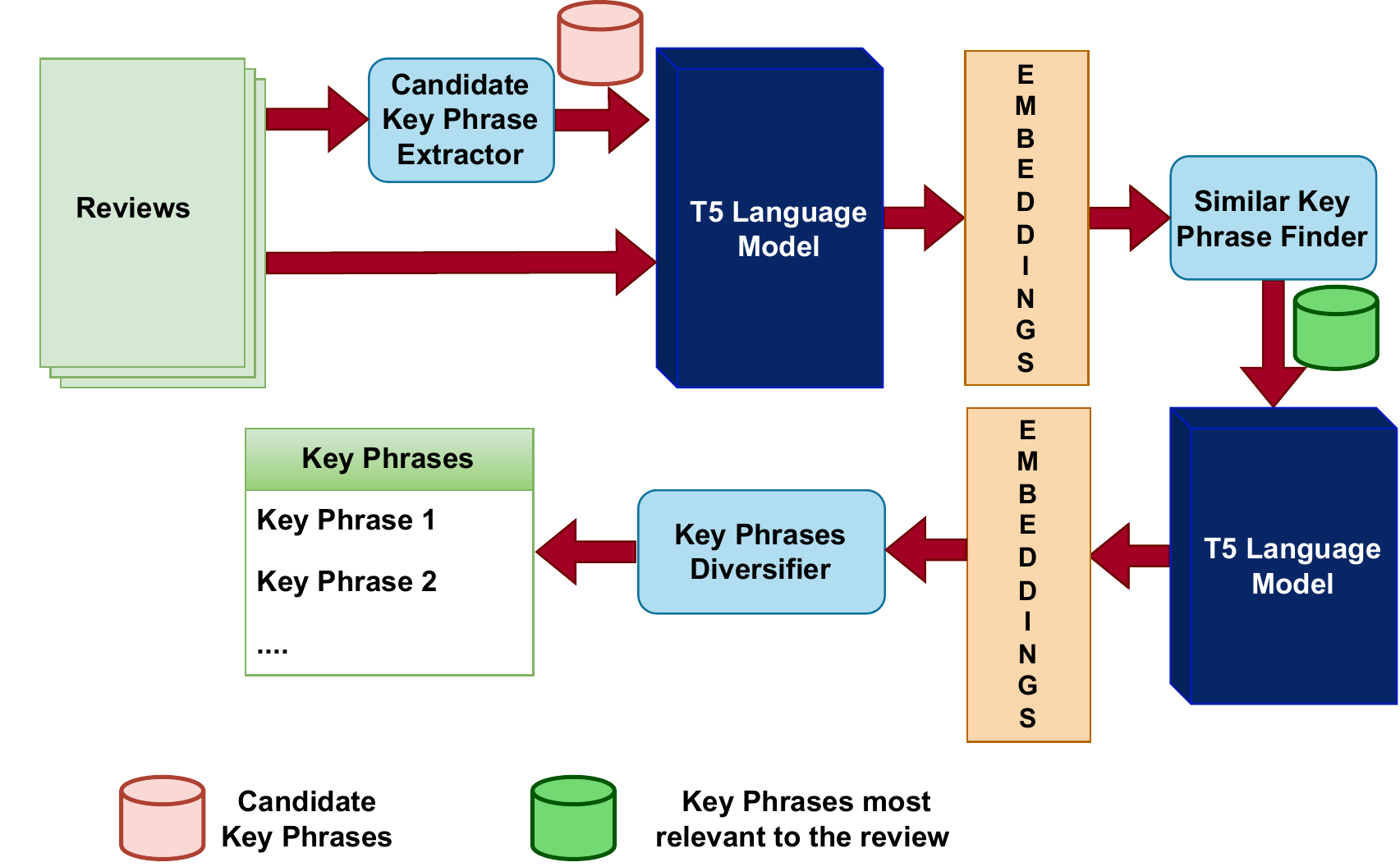}
	
	\label{arc_key_phrase}
\end{figure}
 In Fig.\ref{arc_key_phrase}, we present the proposed Key Phrase Extraction mechanism that is used in the \textit{Review Key Phrase Summarizer} module. In architecture, three main modules are shown.
\begin{itemize}
    \item [-] \textit{Candidate Key Phrase Extractor}: It extracts the key phrases from the reviews based on \textit{n}-gram and Yake~\cite{yake} mechanisms. These key phrases are then fed into T5 model to generate the embeddings. 
    \item[-] \textit{Similar Key Phrase Finder}: It finds the key phrases most relevant to a given review by calculating the semantic similarity score between key phrases and the review based on the embeddings generated by the T5 model.  
    \item [-] \textit{Key Phrase Diversifier}: Among the most relevant key phrases with respect to a particular review, we need to select the most diverse set of key phrases. For that reason, pairwise similarity has been measured among the key phrases based on the embedding generated by the T5 model. In order to ensure the diversity in key phrases, the redundant key phrases with respect to similarity with other key phrases are discarded.  For example - a particular key phrase that is 90\% similar to another key phrase will be discarded in this process.
    
\end{itemize}
\section{Experimental Results}
Enclosed in this annexure is a comprehensive report detailing the experimental outcomes of the framework.
\subsection{Experimental Outcome of Review Sentiment Analyzer Module}
\begin{figure}[htb]
	\centering
	\includegraphics[width=0.49125\textwidth, page=1]{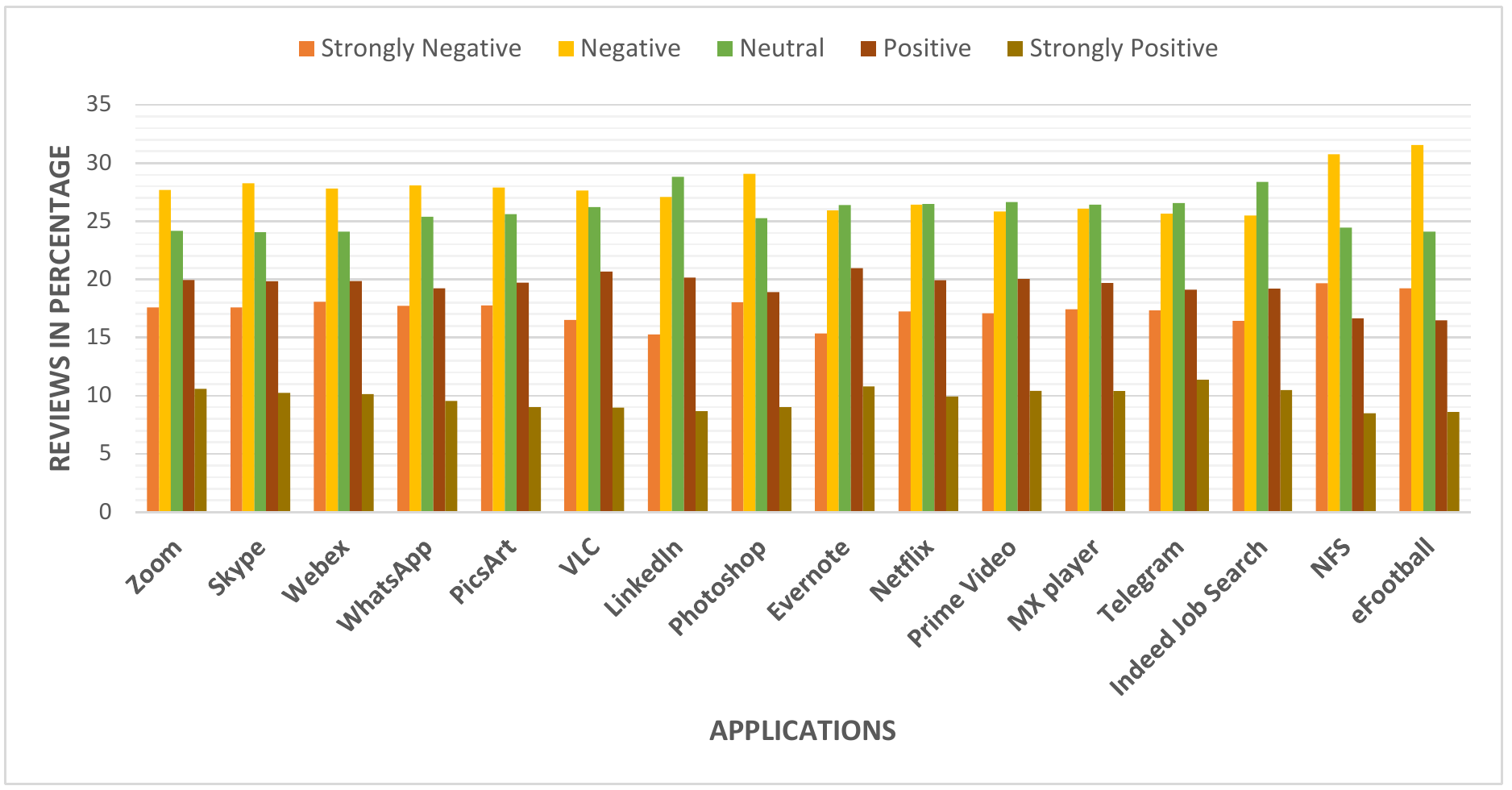}
	\caption{Distribution of reviews (in percentage) for different Apps across 5 class sentiments.
}
	\label{fine-grained-dist}
\end{figure}

The \textit{Sentiment Analyzer} module of the framework recognizes the sentiments of app reviews. The distribution of reviews across the five(5) classes of sentiments is presented in Fig. \ref{fine-grained-dist}. 
\subsection{Experimental Outcome of Review Topic Classification Module}
\begin{figure}[htb]
	\centering
	\includegraphics[width=0.49125\textwidth, page=1]{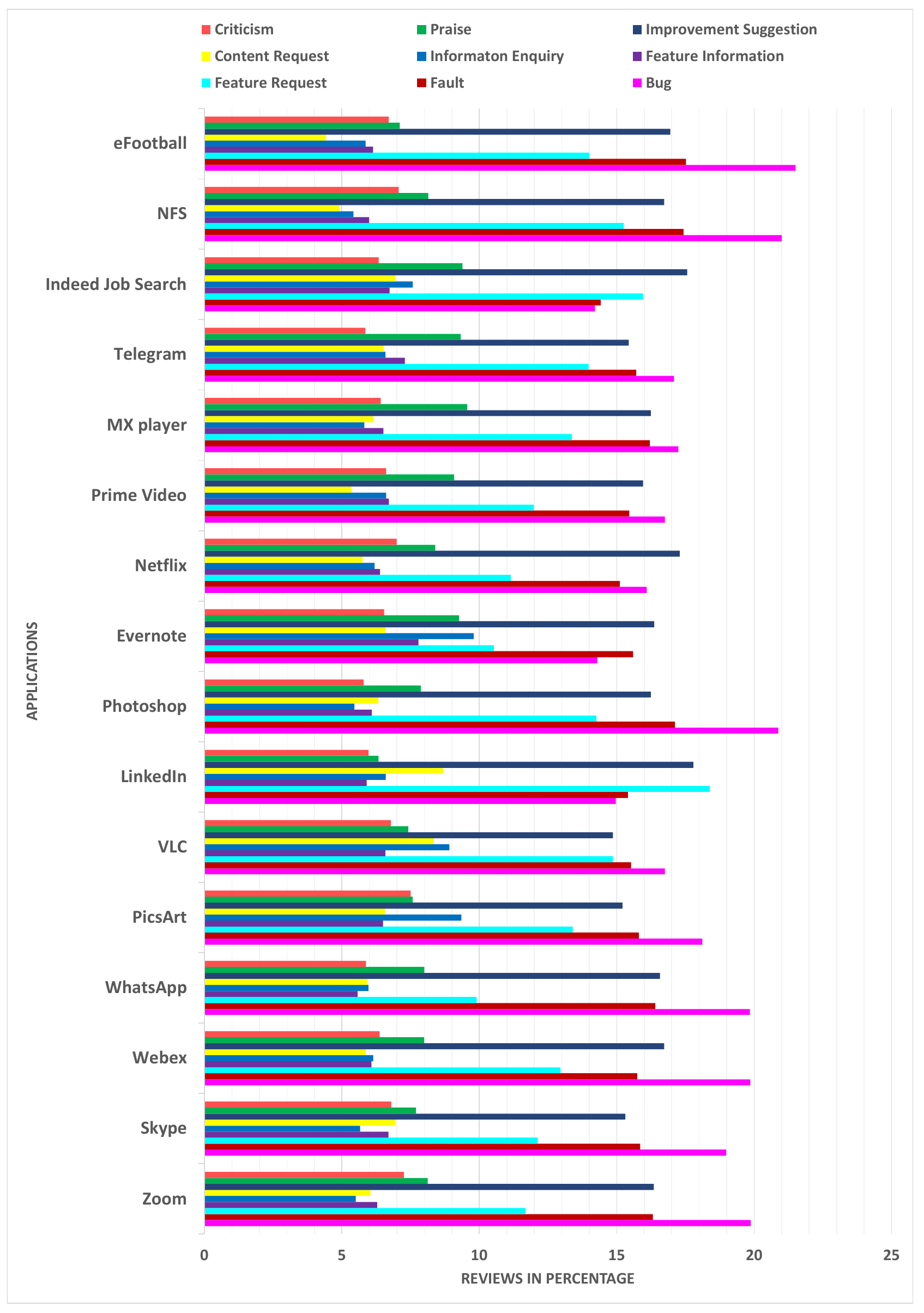}
	\caption{Distribution of reviews across different Topics}
	\label{fig_topic_class}
\end{figure}

\textit{Review Topic Classification } module classifies the reviews into one of the nine(9) categories of topics. Distribution of reviews, for the 16 apps of our case study, across the nine topics is presented in Fig. \ref{fig_topic_class}. Based on our observation, the reporting of bugs is the most prominent aspect in the reviews of apps. Among all the apps, the e-football android gaming app has received the highest number of bug reports from users. However, for the LinkedIn app, there are more feature requests compared to other apps. Meanwhile, for both Indeed Job Search and LinkedIn apps, users have suggested a higher number of improvements.

\begin{figure}[!htb]
	\subfloat[Bug]{\label{fig:sent-bug}
		\includegraphics[width=.48\textwidth, page=1]{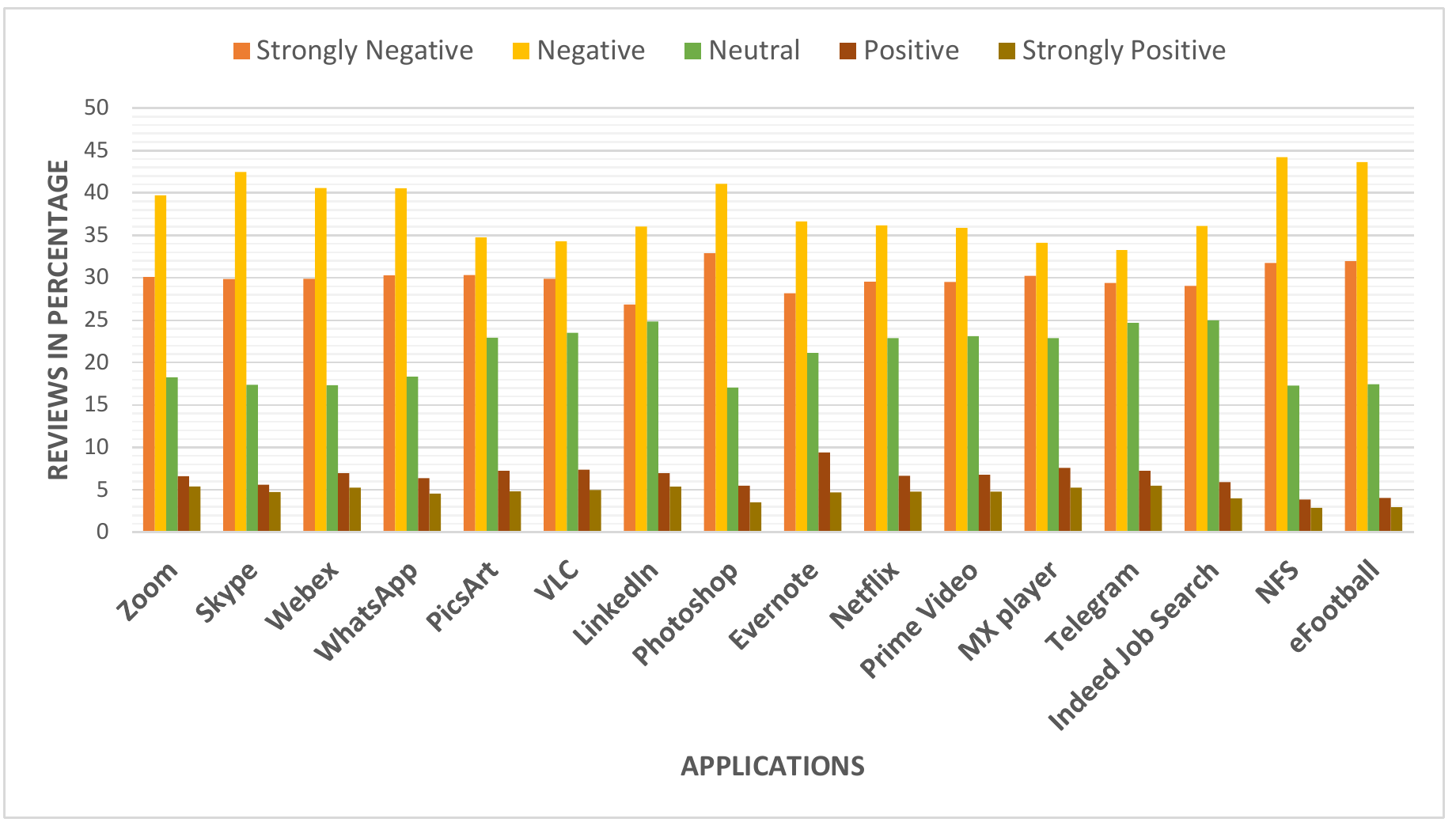}}
		\\[\smallskipamount]
	\subfloat[Feature Request ]{\label{fig:sent-fr}	
		\includegraphics[width=.48\textwidth, page=3]{SentDist.pdf}}
  \\[\smallskipamount]
		\subfloat[Improvement Suggestion]{\label{fig:sent-is}	
		\includegraphics[width=.48\textwidth, page=4]{SentDist.pdf}}
	\\[\smallskipamount]
		\subfloat[Information Enquiry]{\label{fig:sent-ie}	\includegraphics[width=.48\textwidth, page=6]{SentDist.pdf}}
		
	\caption{Distribution of sentiments across different topics.}\label{sent-topic}
\end{figure}

\begin{figure}[!htb]
	\subfloat[Fault]{\label{fig:sent-fault}
		\includegraphics[width=.48\textwidth, page=2]{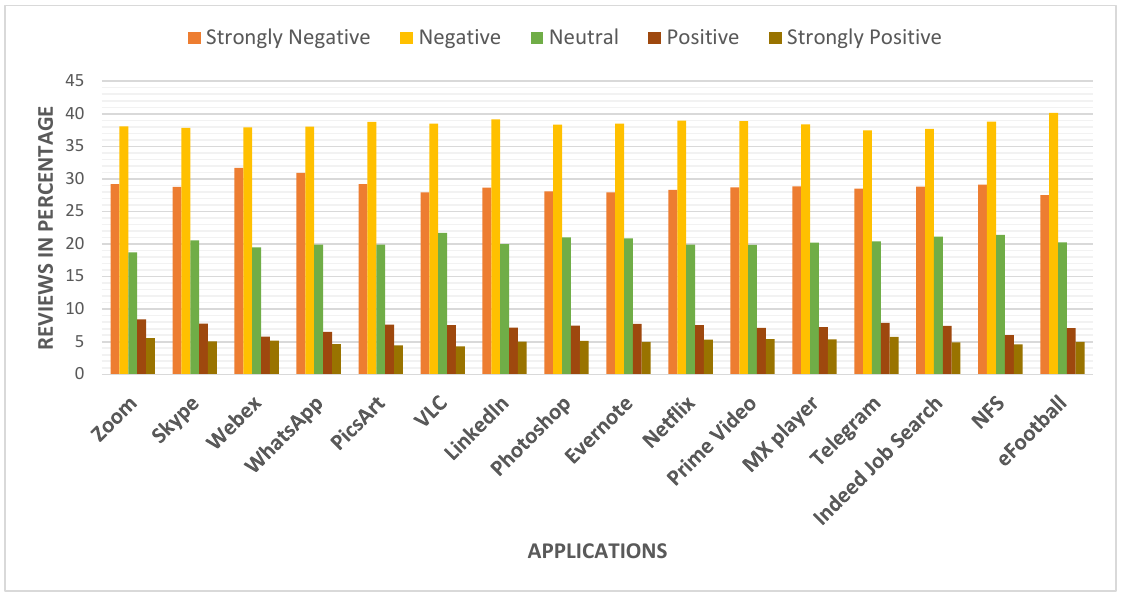}}
		\\[\smallskipamount]
	\subfloat[Feature information ]{\label{fig:sent-fi}	
		\includegraphics[width=.48\textwidth, page=5]{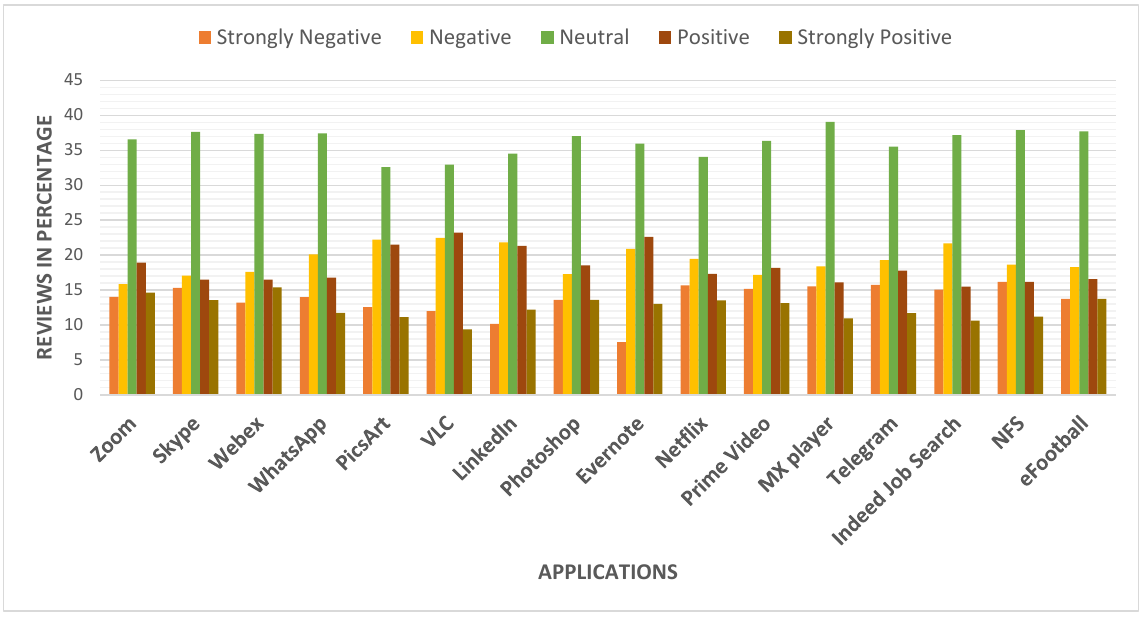}}
  \\[\smallskipamount]
		\subfloat[Content Request]{\label{fig:sent-cr}	
		\includegraphics[width=.48\textwidth, page=7]{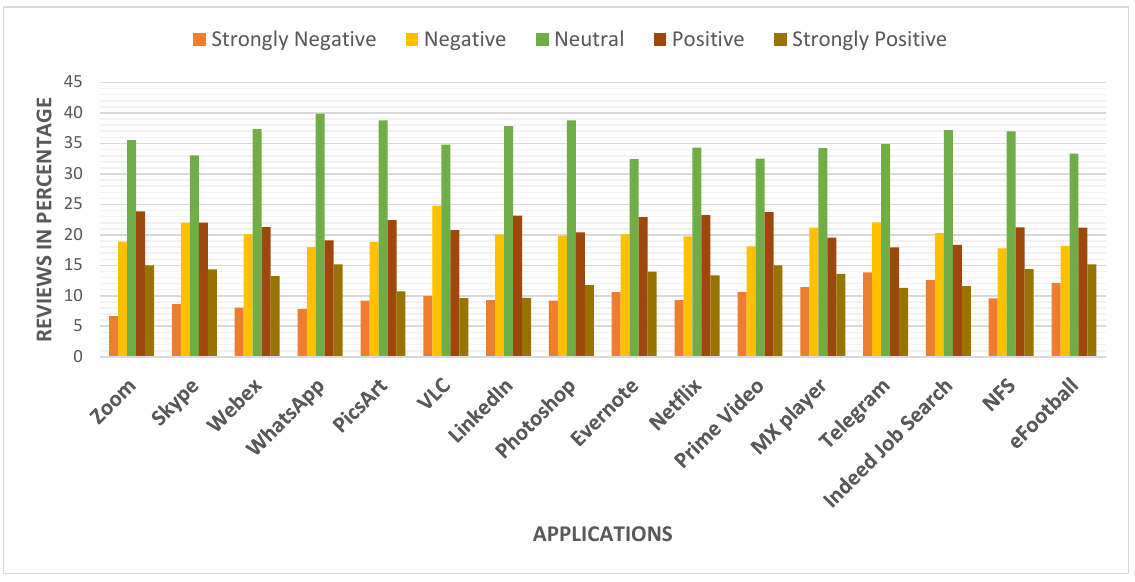}}

	\caption{Distribution of sentiments across different topics.}\label{sent-topic-extended}
\end{figure}

Moving forward, we will examine the distribution of each of the seven topic classes of reviews depicted in Figure \ref{fig_topic_class} across the five classes of sentiments. Our observation indicates that the majority of reviews related to \textit{bugs} (as shown in Figure \ref{fig:sent-bug}) and \textit{faults}((as shown in Figure \ref{fig:sent-fault}) have negative sentiments. This suggests that users are dissatisfied with certain features that are causing bugs and faults in the app. Reviews related to \textit{feature request} (as shown in Figure \ref{fig:sent-fr}) and \textit{improvement suggestion} (as shown in Figure \ref{fig:sent-is}) have sentiments that are mostly neutral or negative. This indicates that the app is lacking some features that some users are unhappy with to some extent. Sentiments related to \textit{feature information} (as shown in Figure \ref{fig:sent-fi}), \textit{information enquiry} (as shown in Figure \ref{fig:sent-ie}), and \textit{content request} (as shown in Figure \ref{fig:sent-cr}) are mainly centered around neutral sentiments. This suggests that users' sentiments regarding these aspects are balanced.

\subsection{Experimental Evaluation of Topic Feature Mapper Module} 
In this section, we consider the user reviews and conduct a semantic search on the corresponding app documentation to identify which specific app features are being discussed in the reviews. This approach enables the developer to gain an understanding of which app features are of concern to users. Next, we present the experimental results of this module for five different apps: Zoom, Skype, Webex, Evernote, and Netflix. For the purpose of brevity, we keep the results of these five apps in the document. 
\subsubsection{Evaluation on Zoom App Reviews}
\begin{figure}[tb]
	\centering
	\includegraphics[width=0.49125\textwidth, page=1]{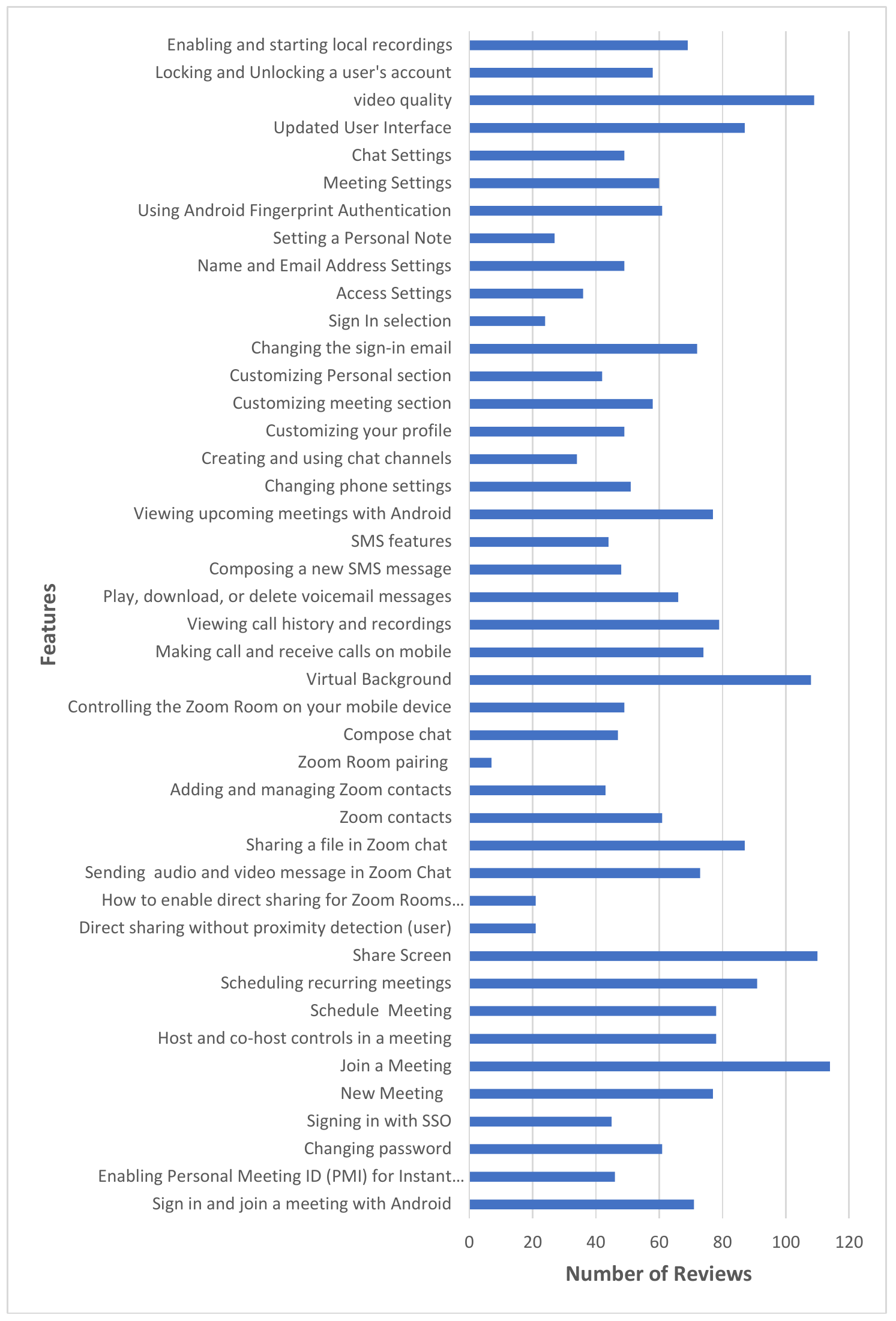}
	\caption{Mapping of Reviews with App Features of Zoom}
	\label{mapping1}
\end{figure}

Figure \ref{mapping1} displays the distribution of reviews across 44 distinct features of the Zoom Android application. Based on the figure, we can identify that several frequently reported issues include \textit{Join a Meeting, Sharing Screen, Sharing a file, Virtual Background, Video Quality, User Interface,} and \textit{Scheduling recurring meetings}.
\subsubsection{Evaluation on Skype App Reviews}
\begin{figure}[!htb]
	\centering
	\includegraphics[width=0.49\textwidth,page=1]{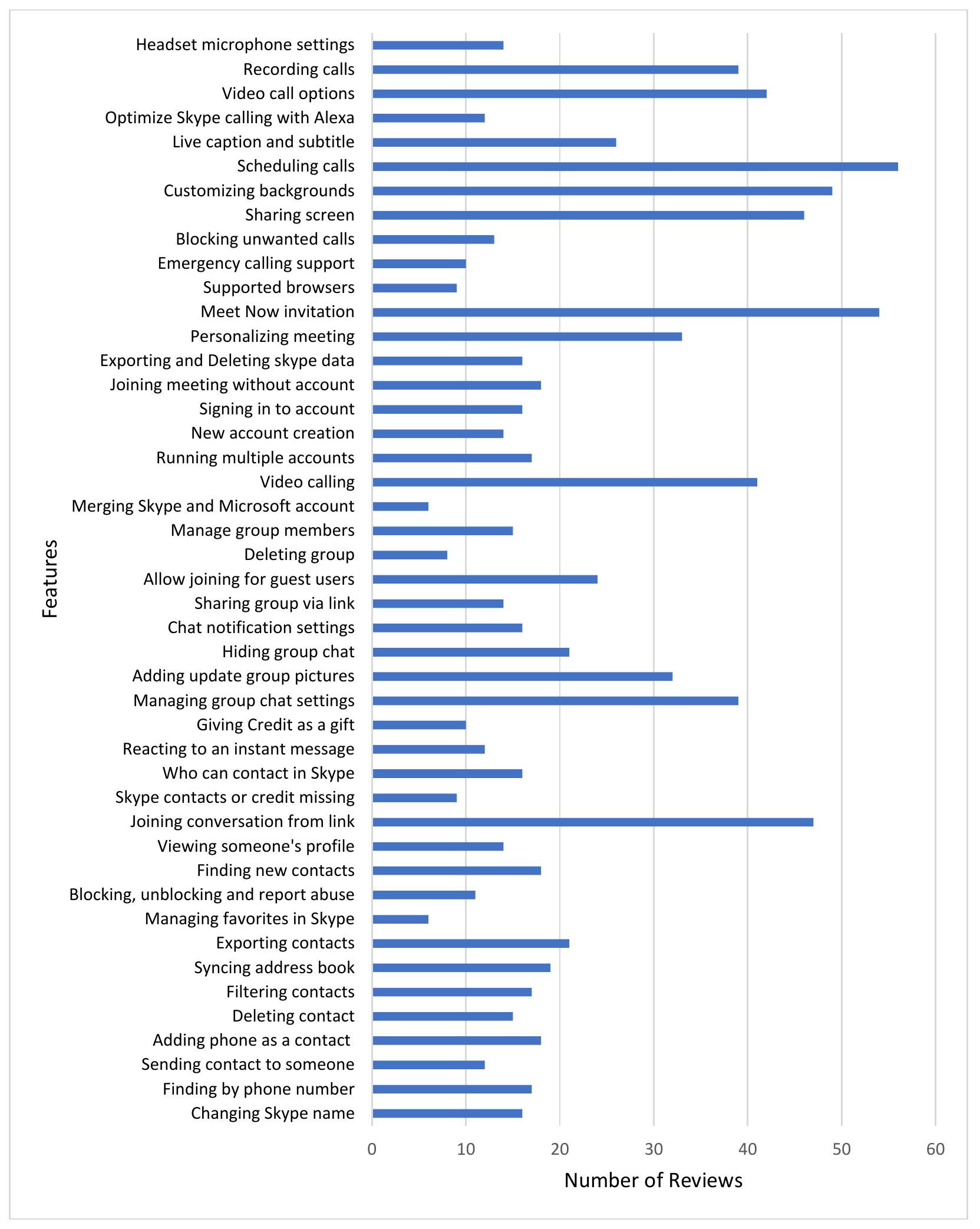}
	\caption{Mapping of Reviews with App Features of Skype}
	\label{mapping2}
\end{figure}

The Fig. \ref{mapping2} shows the distribution of reviews across 47 different features of the Skype Android app. We clearly observe that some of the frequently reported issues include \textit{Scheduling calls, Customizing background, Screen sharing, Meet Now invitation, and Joining conversation from a link}.
\begin{figure}[htb]
\caption{Mapping of Reviews with App Features of Webex}
	\centering
	\includegraphics[width=0.5\textwidth,page=1]{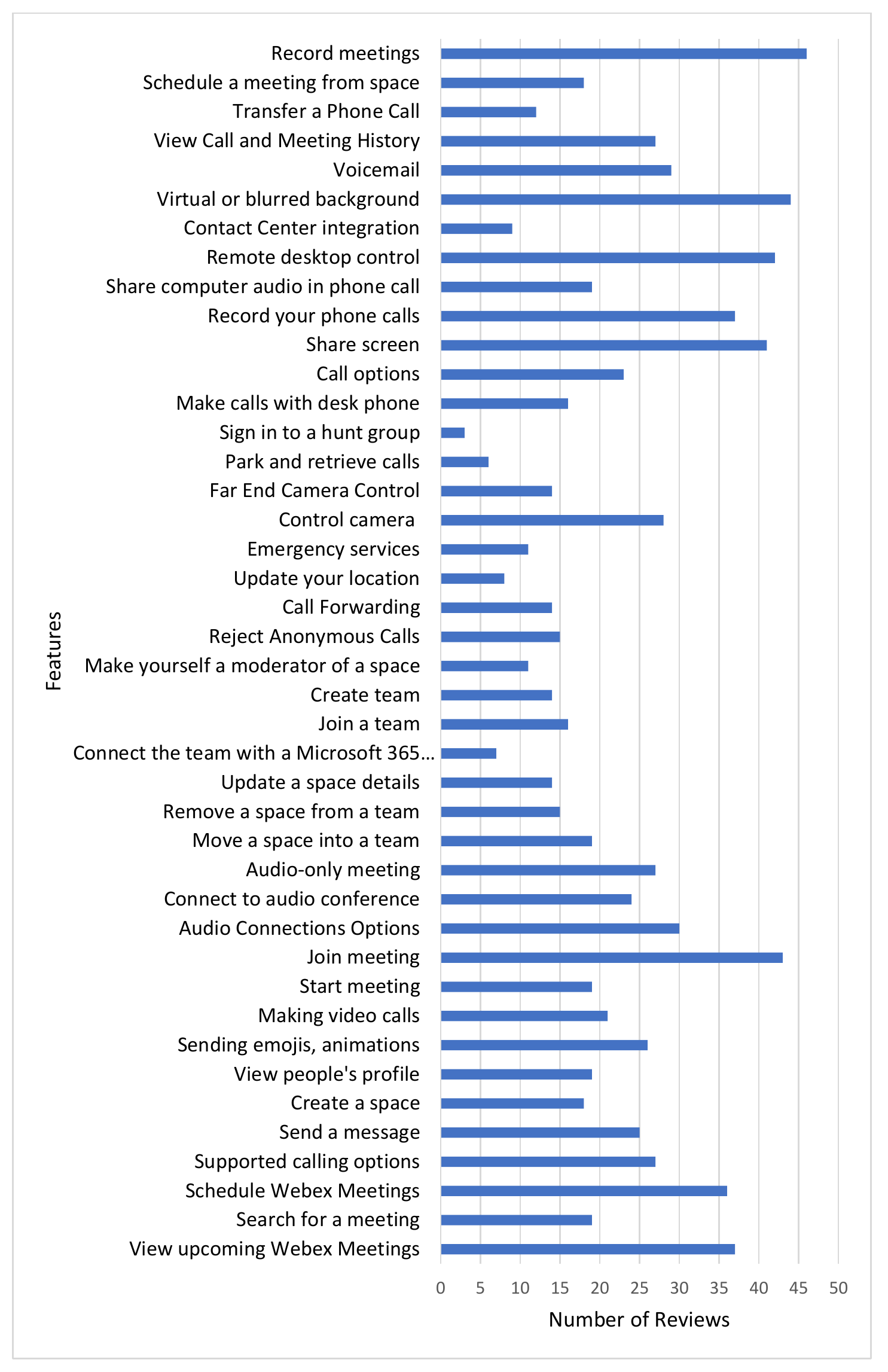}
	
	\label{mapping3}
\end{figure}
\begin{figure}[htb]
\caption{Mapping of Reviews with App Features of Evernote}
	\centering
	\includegraphics[width=0.5\textwidth,page=1]{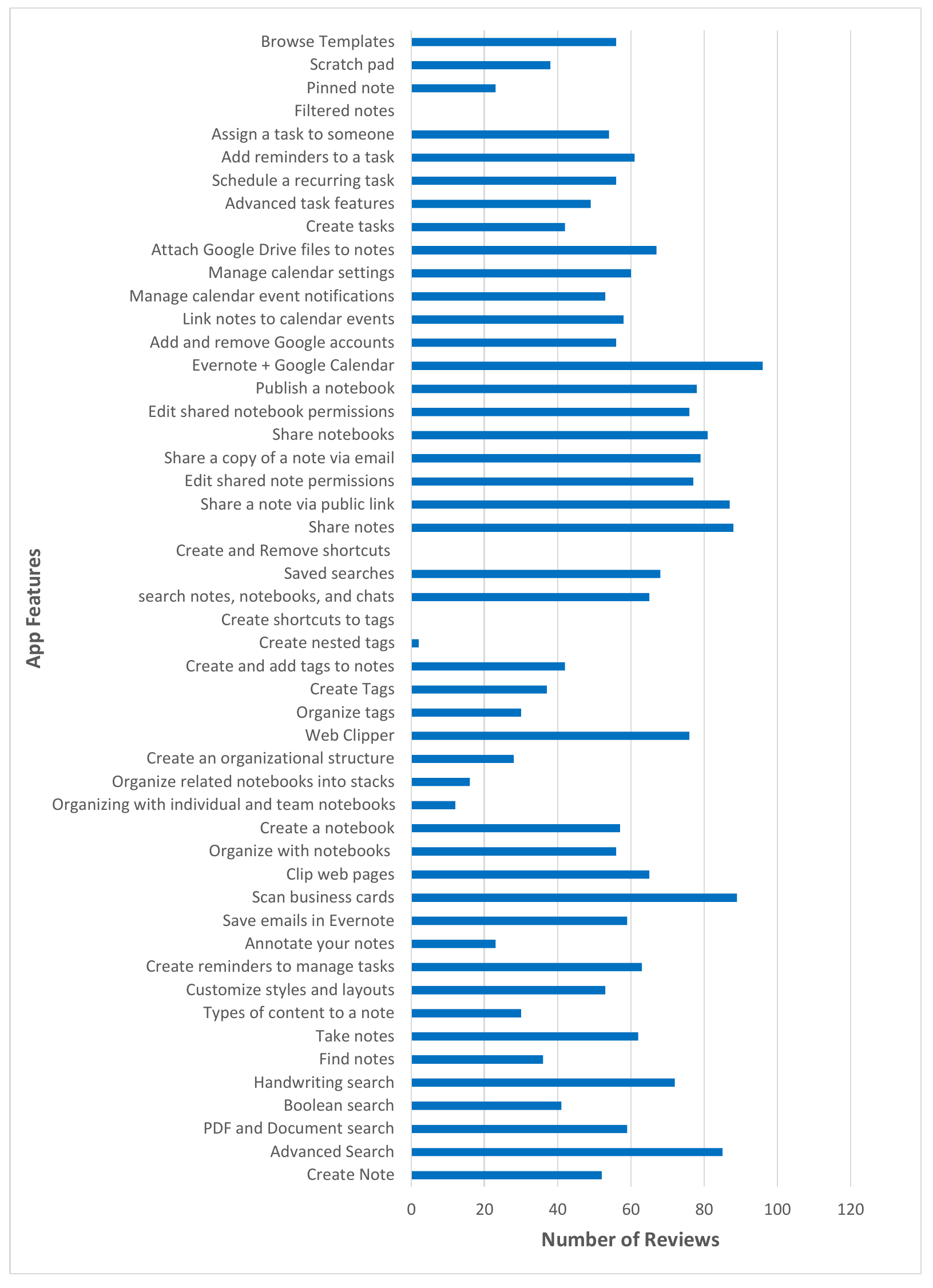}
	
	\label{mapping4}
\end{figure}
\begin{figure}[htb]
\caption{Mapping of Reviews with App Features of Netflix}
	\centering
	\includegraphics[width=0.5\textwidth,page=1]{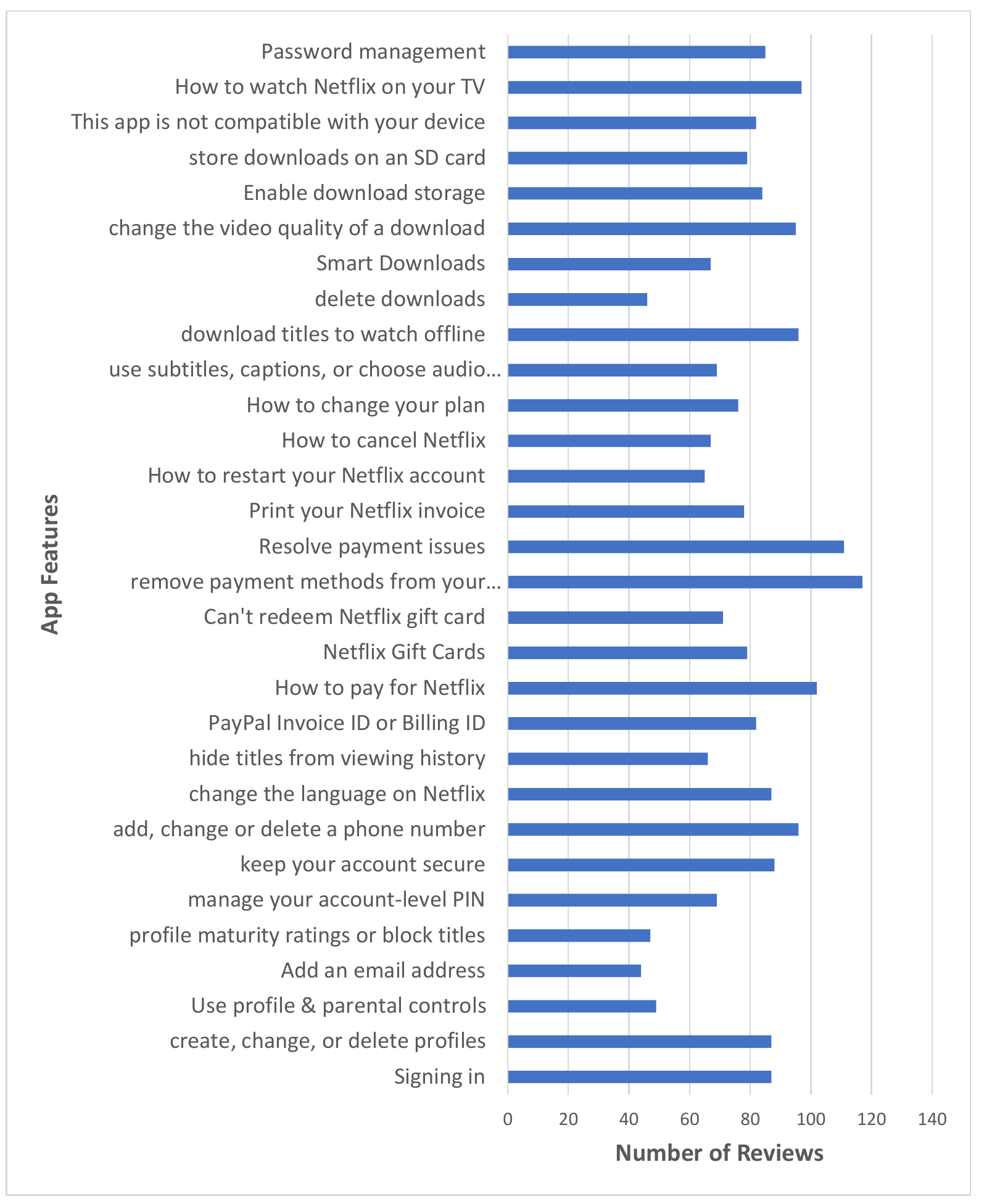}
	
	\label{mapping5}
\end{figure}
\subsubsection{Evaluation on Webex App Reviews}
We evaluate \textit{Topic Feature Mapper} module on reviews of Webex Android application. The mapping of reviews to 42 different features of Webex Android App has been presented in Fig. \ref{mapping3}. It is observable that features like \textit{Record meetings, Virtual Background, Remote desktop control, Join a meeting, sharing screen} have gained attention of the customers.   

\subsubsection{Evaluation on Evernote App Reviews} 
\par  We evaluate \textit{Topic Feature Mapper} module on reviews of the Evernote Android application. The mapping of reviews to 50 different features of the Evernote Android App has been presented in Fig. \ref{mapping4}. It is observable that features like \textit{Evernote and Google Calendar collaboration, scanning business card, advanced search of notes, sharing a note,} and \textit{Sharing note via a public link, sharing screen} have gained the attention of the customers. 
\subsection{Experimental Evaluation of Topic Feature Mapper Module for Netflix App Reviews}
\par  We evaluate \textit{Topic Feature Mapper} module on reviews of the Netflix Android application. The mapping of reviews to 30 different feature-related documentation of the Netflix App has been presented in Fig. \ref{mapping5}. It is observable that documentation related to \textit{ Payments, Watching Netflix on TV} and \textit{Changing phone number} have gained the attention of the customers.

\section{ Effectiveness Evaluation of the Framework}
This section aims to access the effectiveness of our proposed framework through an analysis of a real-world test case scenario. The main objective is to determine whether the framework can efficiently analyze reviews and accurately extract insights from the reviewers. Additionally, it is important to observe that although the framework has not been deployed in the enterprise environment, how well it is still able to capture a significant amount of factual information about the software changes from the app reviews. In order to create such a scenario, we obtained the software release log and app reviews for a period of one year, specifically from December 2021 to February 2023 for the release log and September 2021 to January 2023 for the reviews. Our aim was to determine if the framework could accurately associate reviews with the software release log entries and whether the identified topics align with the actual categories of software change log entries.

To conduct our analysis, we selected the Zoom App as our use case and manually curated 200 release log entries from the app's official website for a one-year period. These entries fell into three categories: Enhancement, New Features, and Resolved Issues. Additionally, we collected approximately 2700 reviews for the same period and used them to conduct our experiment. Using our framework, we obtained a summary of each review and the corresponding topic. We then searched for semantically similar release log entries with each review. To reduce human efforts, we conducted a semantic search and set a threshold of 0.8 for the semantic score to qualify for manual evaluation. This approach helped us identify which reviews were most likely addressed by the development team and reflected in the release log. 

Finally, we evaluated the correspondence between the associated topics of the reviews and the categories of the release log. This manual evaluation allowed us to determine the effectiveness of our framework in accurately identifying relevant information and associating it with the appropriate software changes.
Our analysis, presented in Table \ref{tbl_validation}, shows that a significant percentage of the release log entries for Enhancement, New Features, and Resolved Issues were addressed in the app reviews, regardless of the alignment between the review topic and the release log category. Specifically, we found that 48\%, 55\%, and 57\% of the entries for Enhancement, New Features, and Resolved Issues respectively were successfully attended to in the app reviews.
\begin{table}[htb]
\caption{Mapping of reviews to release log entries with respect to different categories}
\resizebox{.5\textwidth}{!}{\begin{tabular}{|l|c|c|}
\hline
\textbf{Categories} & \multicolumn{1}{l|}{\textbf{\begin{tabular}[c]{@{}l@{}}Percentage of Release log   \\ entries mentioned in Reviews \\ correctly\end{tabular}}} & \multicolumn{1}{l|}{\textbf{\begin{tabular}[c]{@{}l@{}}Percentage of Release log   \\ entries mentioned in Reviews \\ irrespective of the categories\end{tabular}}} \\ \hline
Enhancement         & 40                                                                                                                                             & 48.23529412                                                                                                                                                         \\ \hline
New Features        & 38.88888889                                                                                                                                    & 55.55555556                                                                                                                                                         \\ \hline
Resolved Issues     & 41.02564103                                                                                                                                    & 57.69230769                                                                                                                                                         \\ \hline
\end{tabular}}
\label{tbl_validation}
\end{table}
We also observed that a considerable percentage of the release log entries were correctly identified in the app reviews with regard to the one-to-one correspondence between the review topic and the release log category. Notably, 40\%, 38\%, and 41\% of the entries for Enhancement, New Features, and Resolved Issues respectively were accurately recognized in the app reviews.
However, it is worth mentioning that during the semantic search filtering process before manual evaluation, we may lose some of the reviews that are genuinely associated with the release log entries. Hence, the actual number of correctly identified release log entries in the app reviews could be slightly higher than what is presented in the table.

Overall, this analysis demonstrates the potential of our framework to provide valuable insights to development teams and improve the software development process. By accurately identifying user feedback and associating it with relevant software changes, our framework can help developers understand how their software is being received and identify areas for improvement.

\end{document}